\documentclass{ws-procs975x65}
\usepackage{wrapfig}

\def\journal#1#2#3#4{ {#1} {\bf #2}, {#3}\  ({#4})}
\def\AmJPhys{\journal{Am.\ J.\ Phys.}}

\def\JMathPhys{\journal{J.\ Math.\ Phys.}}

\def\MPLA{\journal{Mod.\ Phys \ Lett. \ {\bf A}}}

\def\NPProc{\journal{Nucl.\ Phys.\ Proc.\ Suppl}}
\def\NuclPhys{\journal{Nucl.\ Phys.}}

\def\PLB{\journal{Phys.\ Lett.\ {\bf B}}}
\def\PhysRev{\journal{Phys.\ Rev.}}

\def\PRD{\journal{Phys.\ Rev.\ {\bf D}}}

\begin{document}

\title{The Elusive $\nu$ mass since 1933}

\author{Ngee-Pong Chang}

\address{The City College \& The Graduate Center of \\
The City University of New York \\
New York, NY 10031\\
$^*$E-mail: npccc@sci.ccny.cuny.edu\\
%www.university\_name.edu
}

\begin{abstract}
We review briefly the history of the $\nu_e$ mass since 1933, and point out how the KATRIN experiment can resolve the mystery of the tachyonic mass indicated by the Mainz-Trotsky experiments.
\end{abstract}


\bodymatter

\section{Solvay Congress 1933}
\begin{wrapfigure}{r}{0.74\textwidth}
\vspace{-.30in}
\begin{center}
\includegraphics[width=0.74\textwidth,angle=0]{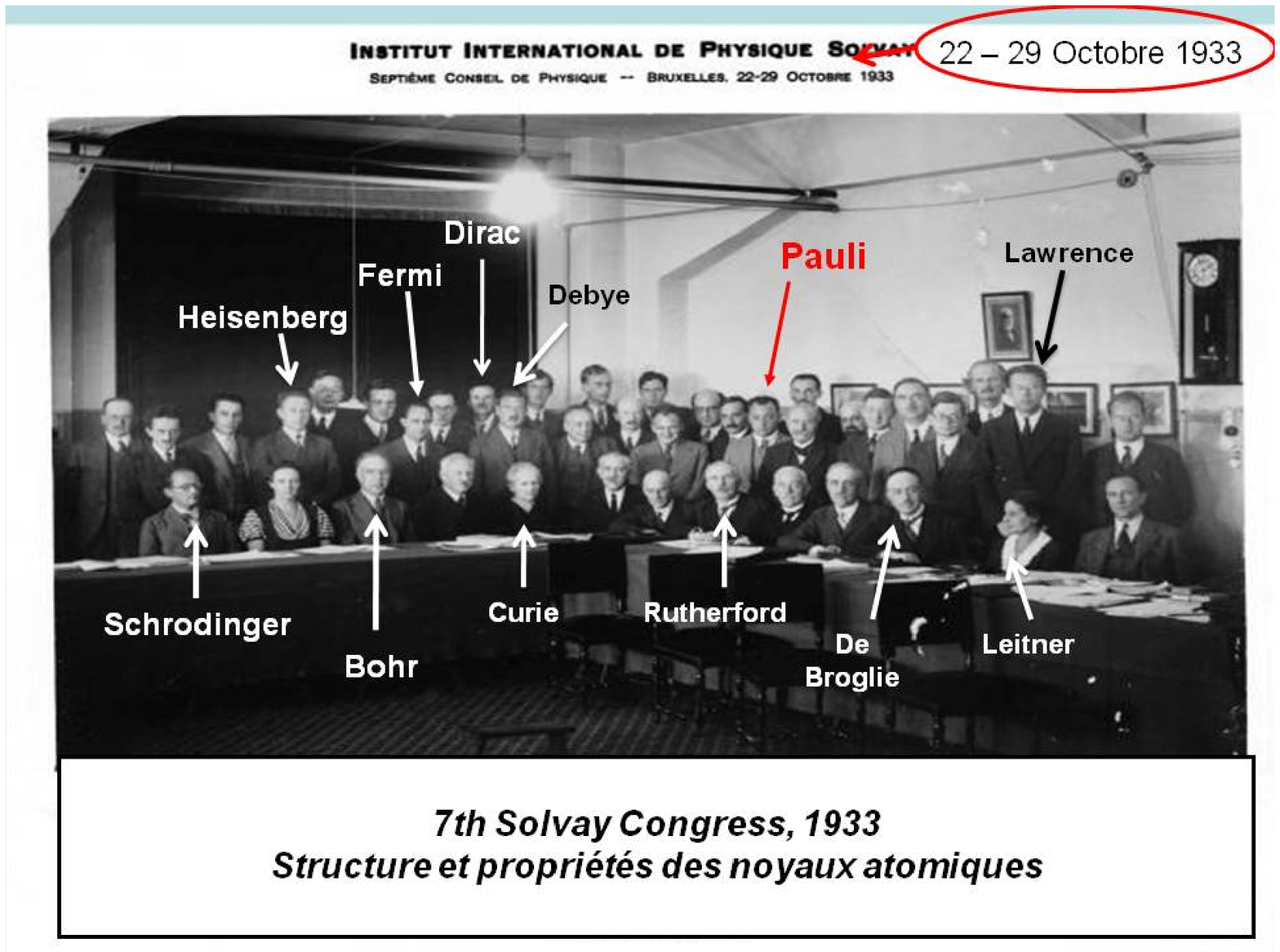}
\end{center}
\vspace{-.3in}
\end{wrapfigure}
Many people have wondered at the title of this conference, Particle \& Astrophysics, Quantum Field Theory: 75 years since Solvay.  The people who have access to Wiki will quickly discover that the first Solvay Congress was in 1911, and this being 2008, the math does not work out.  But the title of the conference actually points to the Seventh Solvay Congress in 1933, and as one of the earlier speakers pointed out, since E.O. Lawrence was at this Congress, particle physics truly was born 75 years ago.

\vspace{.2in}
This Solvay Congress '33 was notable for the offical birth of the neutrino and the effective $4-$fermion field theory.  The story of the neutrino goes back to the famous postcard (dated 4 Dec, 1930, and addressed to `Radioactive Ladies and Gentlemen') which Pauli wrote to the physics meeting at T$\ddot{u}$bingen.  In this postcard, Pauli wrote:  "`{\em I have hit upon a desperate remedy to save the 'exchange theorem' of statistics and the law of conservation of energy.  
Namely, the possibility that there could exist $\ldots$  electrically neutral particles, that I wish to call \underline{neutrons}, which have spin 1/2 and obey the exclusion principle $\ldots$ The continuous beta spectrum would then become understandable by the assumption that in beta decay a \underline{neutron} is emitted in addition to the electron such that the sum of the energies of the neutron and the electron is constant...    }"'(http://wwwlapp.in2p3.fr/neutrinos/anhistory.html
)
%http://wwwlapp.in2p3.fr/neutrinos/anhistory.html
\begin{wrapfigure}{l}{0.4\textwidth}
\vspace{-.1in}
\begin{center}
\includegraphics[width=0.19\textwidth,angle=0]{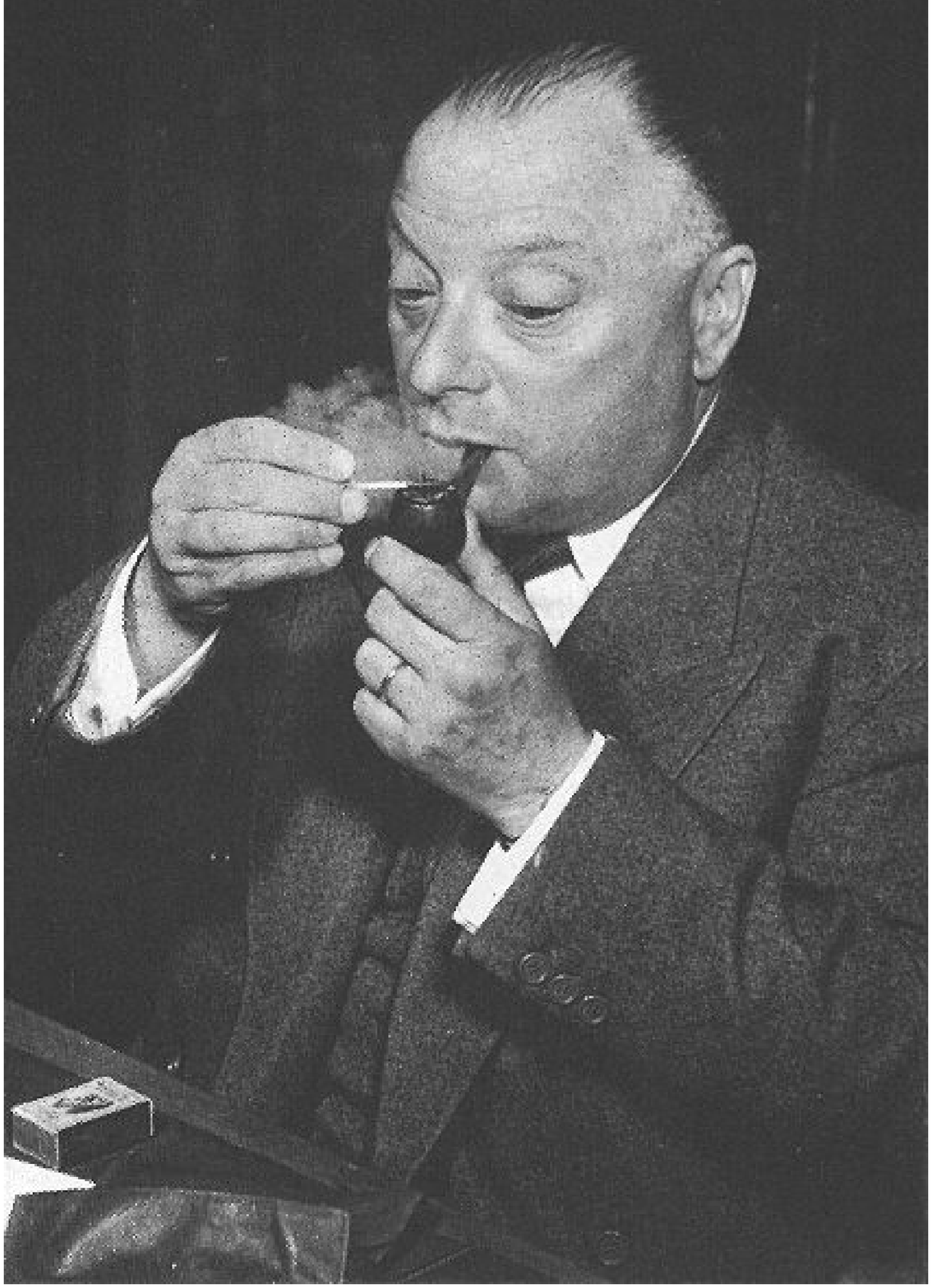}
\noindent \hrulefill
\includegraphics[width=0.17\textwidth,angle=0]{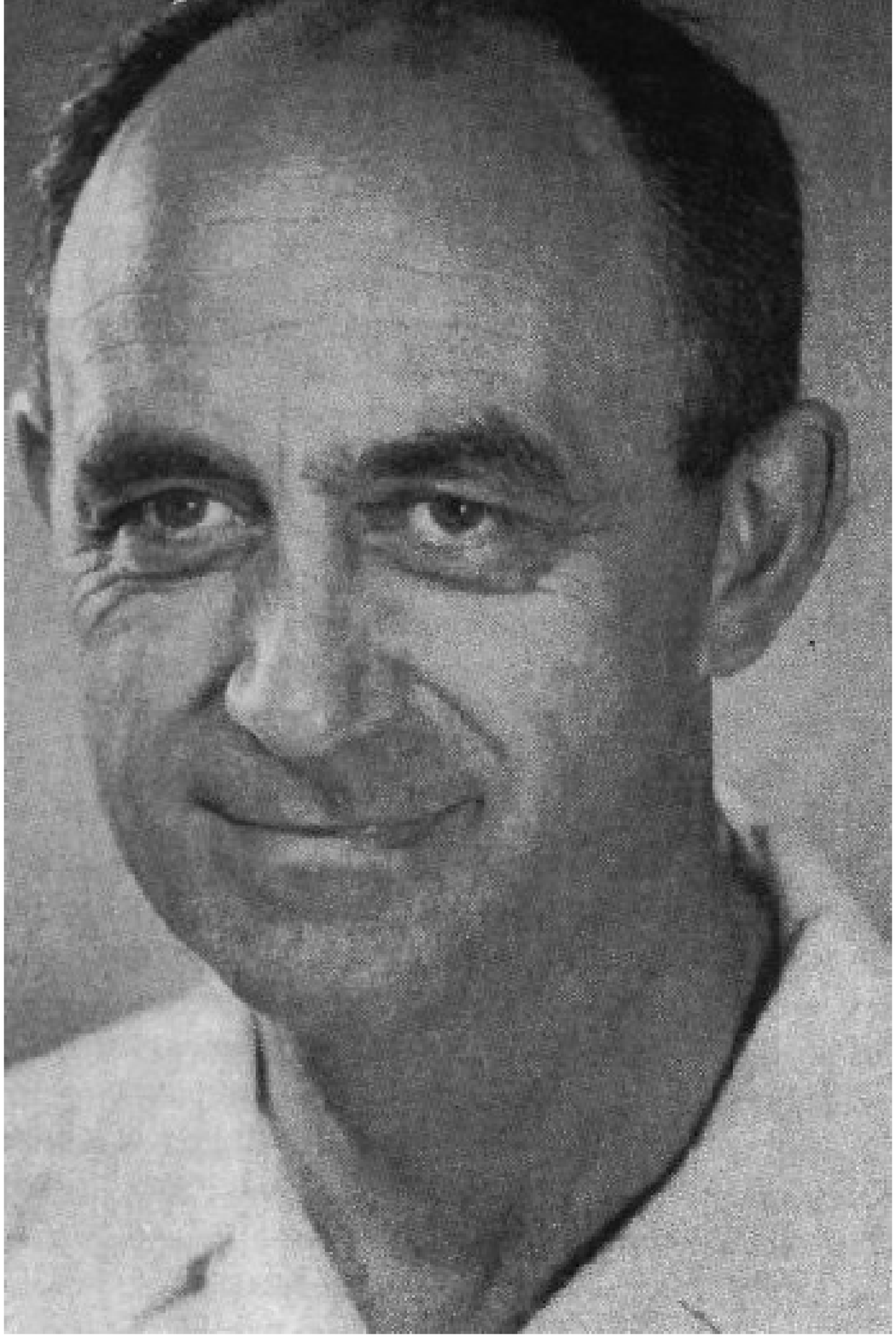}
\label{fig-Pauli-Fermi}
\caption{(a) Pauli  \hspace*{20pt} (b) Fermi }
\end{center}
\vspace{-.0in}
\end{wrapfigure}

This was indeed a bold idea, but the name was not correct, because Chadwick discovered the real neutron in February 1932.  At Fermi's suggestion, Pauli renamed his particle \underline{neutrino} (the little neutral one), and gave a talk on his proposal at the Solvay '33.  
\begin{wrapfigure}{c}{0.4\textwidth}
\vspace{-.3in}
\begin{center}
\includegraphics[width=0.4\textwidth,angle=0]{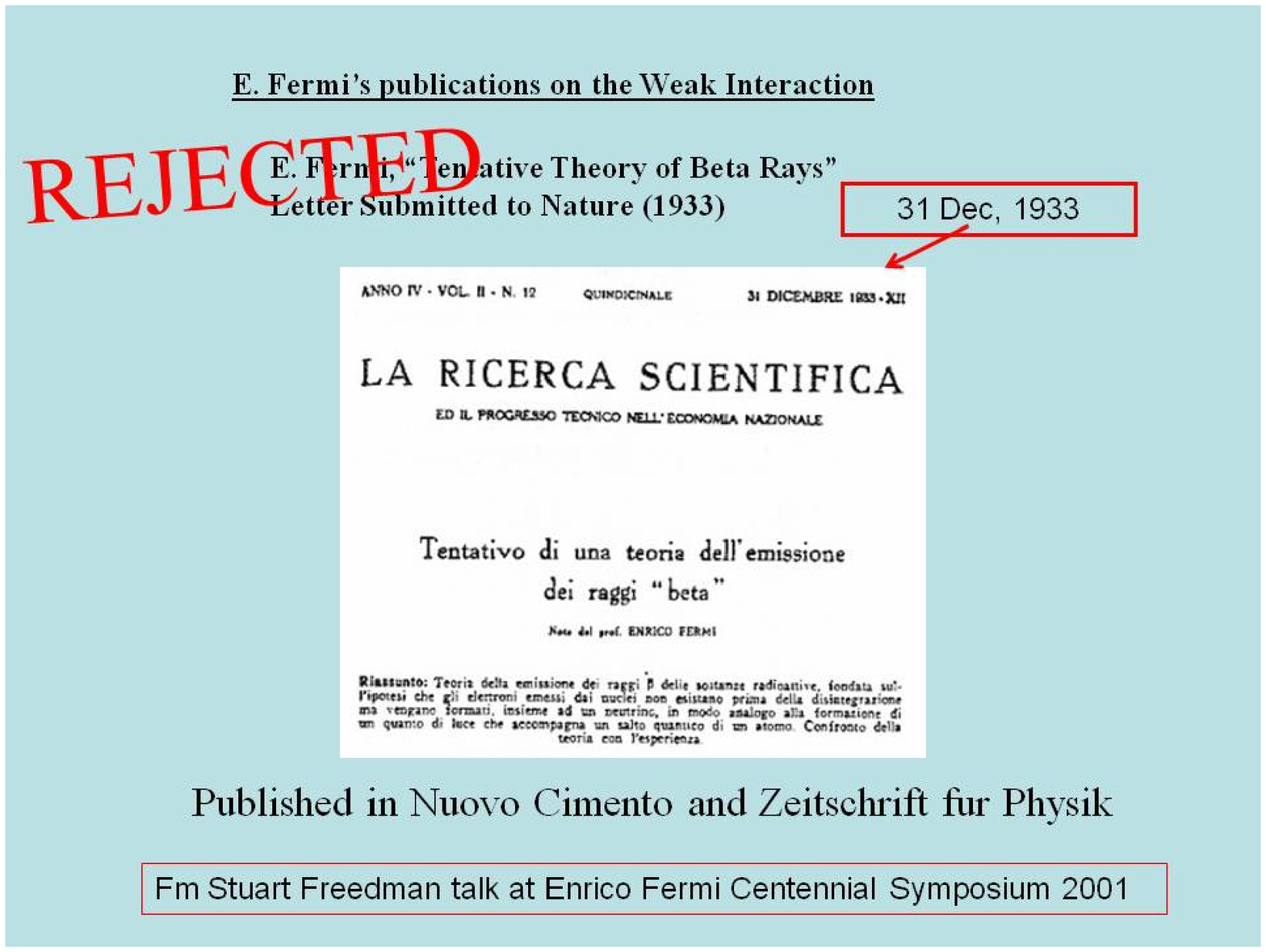}
\label{fig-Fermi_paper}
\end{center}
\vspace{-.3in}
\end{wrapfigure}
Fermi was at the Solvay Congress.  After the congress, he quickly wrote up a paper "`Tentative Theory of Beta Rays"', submitted it to Nature (1933), but was rejected.  Fortunately, rather than be dejected, he quickly submitted it to Nuovo Cimento, and the paper was published 31 Dec, 1933.  In it, as well as a follow-up paper in Zeitschrift der Physik, the famous $4-$fermion effective Lagrangian for $\beta-$decay was written down.(S Freedman lecture Fermi Centennial Symposium 2001)

\vspace{.2in}
Already in 1933, F. Perrin observed that the neutrino mass had to be very much lower than the electron mass. (Comptes Rendus 197 (1933) 1625)

\section{Mainz-Troitsk result}

\begin{wrapfigure}{r}{0.5\textwidth}
\vspace{-.15in}
\begin{center}
\includegraphics[width=0.5\textwidth,angle=0]{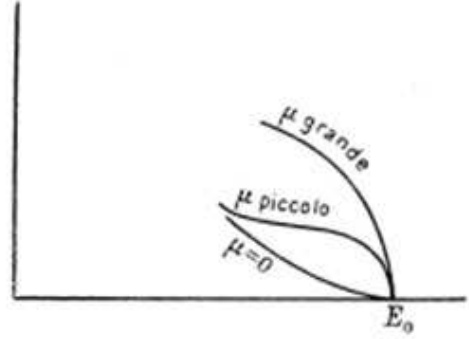}
\vspace{-.25in}
\label{fig-Fermi_endpoint}
\caption{Fermi sketch end-point scenario}
%\subfigure[Large]
\end{center}
%\end{figure}
\vspace{-.2in}
\end{wrapfigure}
In the intervening years, there have been a heroic series of experiments devoted to the measurement of the electron neutrino mass.  The best method is to measure the end-point behavior of the $\beta-$decay spectrum of tritium
${}^3H \;\longrightarrow\; {}^3 He \;+\; e^{-} \;+\; \nu_e $
By the $4-$Fermion theory, the shape of the $\beta-$decay spectrum is dependent on the $\nu-$mass. Fermi himself sketched the shape of the end-point spectrum and speculated on the `piccolo' vs the `grande' scenario for the neutrino mass.  In the notation preferred by the experimentalists, the spectrum is given by ($p, E$ are the momenta and the total energy of the electron, while ${\cal E}$ refers to the kinetic energy, and $\Delta = M' - M$)
\begin{eqnarray}
	\frac{1}{\tau} &=& \frac{4G_F^2}{\pi^3} \frac{M'}{M} {\displaystyle \int } d{\cal E} \;p E (\Delta - m - {\cal E})
											\sqrt{(\Delta - m - {\cal E})^2 - m_{\nu}^2}	\label{decay-rate}
\end{eqnarray}

Precision measurement of the end-point behavior of the tritium beta-decay spectrum would thus give a direct measurement of the $\nu_{e}$ mass.  While in the early days, the experimental errors did not allow a precise determination, all results point to a small or vanishing mass for the electron neutrino.  It is a tribute to the unwavering determination of the small group of nuclear physics experimentalists that the limit on the electron neutrino mass came down to the $eV$ level.  The latest results reported in 1998-99 were carried out by the Troitsk group in Russia(ref ~\refcite{Troitsk}) and Mainz group in Germany(ref ~\refcite{Mainz}).  Their experiments led to the startling results:
%\vspace{-.1in}
\begin{equation}
%\vspace{.2in}
	\left. {\begin{array}{rclcl}
	m^2_{\nu_e}  &=& ( - 1.6 \pm 2.5 \pm 2.1 ) \;eV^2  && {\rm (Mainz)} \\
	m^2_{\nu_e}  &=& ( - 2.3 \pm 2.5 \pm 2.0 ) \;eV^2  && {\rm (Troitsk)} 
				\end{array}} \right\}
\end{equation}

\def\figsubcap#1{\par\noindent\centering\footnotesize(#1)}
\begin{figure}[hbtp]%
\begin{center}
 \parbox{2.1in}{\epsfig{figure=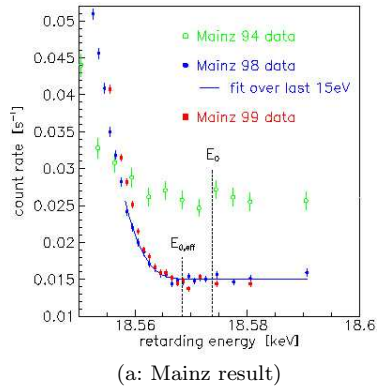,width=2in}
 \figsubcap{a: Mainz result}}
 \hspace*{4pt}
 \parbox{2.1in}{\epsfig{figure=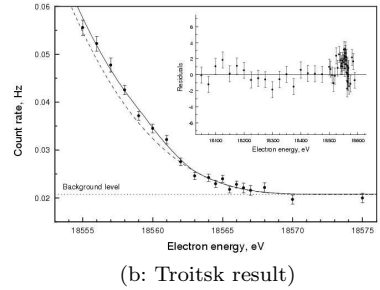,width=2in}
 \figsubcap{b: Troitsk result}}
\label{Mainz-Troitsk}
 \caption{
 Tritium end-point measurements by Mainz and Troitsk groups}
\end{center}
\vspace{-.4in}
\end{figure}

\vspace{.2in}
The shape of the end-point spectrum does not conform to the parabolic shape of a normal mass-squared neutrino.  Instead, the curve appears to fit a negative mass-squared neutrino, what my thesis advisor at Columbia had named in 1978 as tachyon ! (Ref~\refcite{tachyon})

This unexpected result on $m_{\nu}^2$ has led to much soul-searching among both the theorists and the experimentalists.  Among the experimentalists, there is even greater determination to accurately measure the end-point, leading to the formation of the KATRIN experiment now going on at Karlsruhe.  This experiment involves the use of such a giant spectrometer that shipping it from the manufacturer in Deggendorf (outside Munich) to the KATRIN laboratory site in Karlsruhe involved a tortuous detour of about $8800 \;km$ via shipping on the Danube to the Black Sea port of Constanta, and then via container ship around Europe to the North Sea port, and finally down the Rhine to Karlsruhe.  The final transport through the village of Leopoldshafen involved the town police \& fire department escort past the narrow passageway between houses.  I urge all interested readers to go to their website and then you will appreciate the enormity of the logistical undertaking in pursuing fundamental physics. 
\begin{wrapfigure}{r}{0.5\textwidth}
\begin{center}
\includegraphics[width=0.5\textwidth,angle=0]{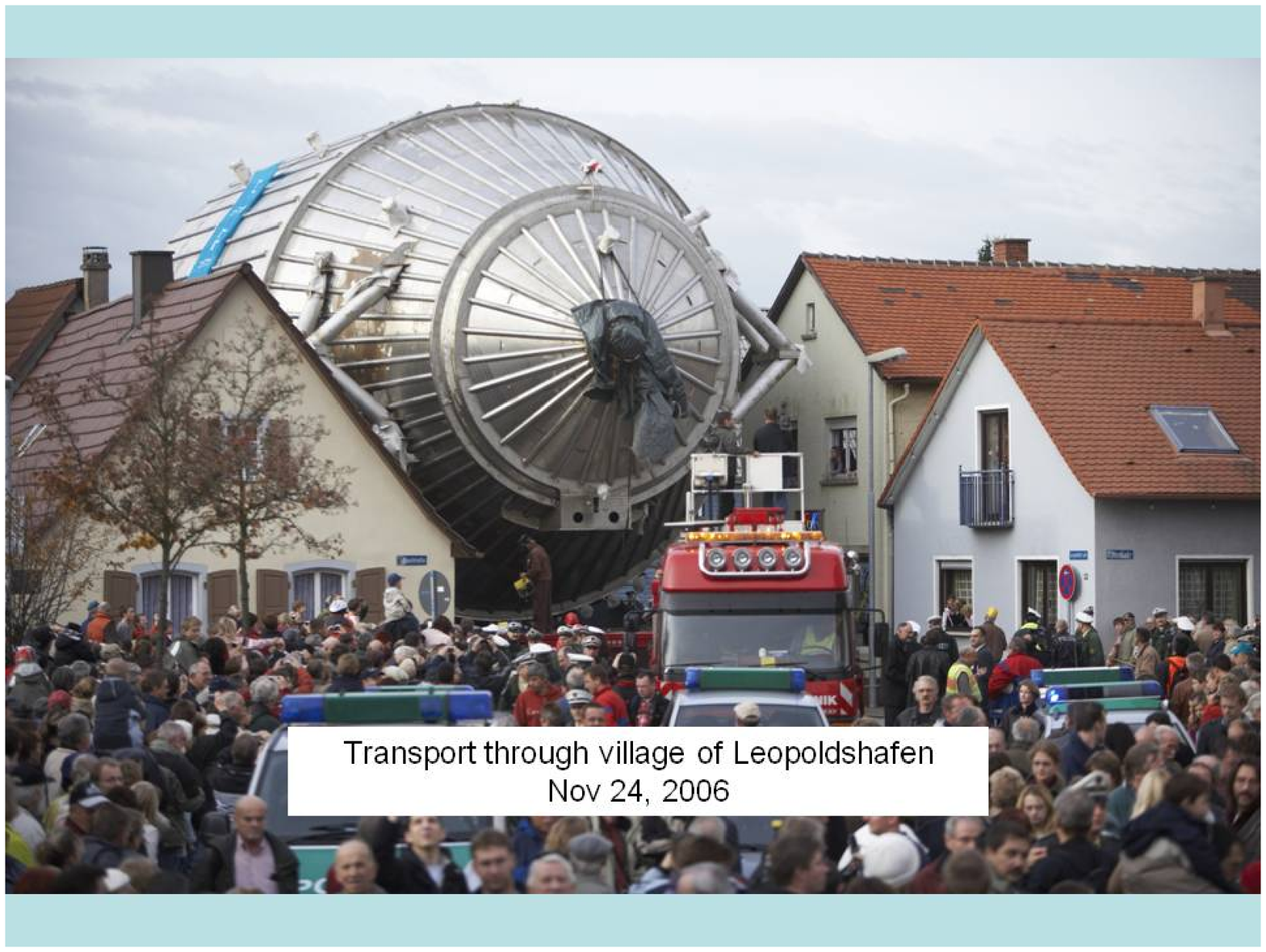}
\end{center}
%\label{fig-Katrin detector}
%\end{figure}
\vspace{-.4in}
\end{wrapfigure}

\vspace{.2in}
The goal of KATRIN is to measure the electron neutrino mass down to an accuracy of $0.2 \;eV$.  They are on target to measure  tritium decays, and are concentrating on the last $30 \;eV$ of the electron kinetic energy.  According to their estimate, the last $5 -15\;eV$ of the decay spectrum corresponds to roughly $10^{-10}$ of the total decay events!

\section{Tachyon Neutrino}

Among theorists, there was the usual divergence of opinions.  In the mainstream, there are those who dismiss the quoted result as premature, and the tachyonic conclusion to be impossible.  Indeed, while the Particle Data Group at first did quote the Mainz-Troitsk result as I have stated it above, after some time, they relegated that result to a footnote, and instead cast their result as being consistent with $m_{\nu_{e}} < 2.3 \;eV$.
In the minority are those theorists willing to consider the impossible, in the same spirit as Pauli, Fermi, Bohr and others did in the 1930's.  In 1985, based on the data already indicated then by ITEP and others, Chodos, Hauser and Kostelecky (ref ~\refcite{tachyon-chodos}) suggested that a tachyonic neutrino would obey the pseudo-Dirac equation
\begin{eqnarray}
	i \gamma \cdot \partial \gamma^5 \;\Psi  &=& m \Psi     \label{pseudo-Dirac}
\end{eqnarray}
And indeed, in a paper in 2001, I worked on this Chodos-Hauser-Kostelecky model of the neutrino, and showed how the canonical quantization of this tachyonic field theory can be carried out, with full 
micro-causality\footnote{\samepage
The $\gamma^5$ in the equal-time anti-commutator rule implies the presence of a negative metric right-handed neutrino component.  Because of the $V-A$ nature of the weak interaction, this negative metric component remains sterile.  Instead, its role is in forming a Nambu-Jona-Lasinio $in$ and $out$ vacua.  See ref ~\refcite{Chang}.
}
\begin{eqnarray}
	\{ \Psi_{\nu} (\vec{x}, x_o), \Psi_{\nu}^{\dagger}(\vec{y}, y_o)  \}{\displaystyle |}_{x_o = y_o} &=& \gamma^5 \delta (\vec{x} - \vec{y})  			\label{equal-time}
\end{eqnarray}
The key to that quantization is to recognize that the Fourier components of the tachyonic field consist not only of the mass-shell tachyonic momenta with $\vec{q}\,{}^2 > m_{\nu}^2$, but also the transient solutions of the pseudo-Dirac equation, where $\vec{q}\,{}^2 < m_{\nu}^2$.  These latter modes lead to the exponential decay and run-away solutions.  The canonical quantization shows that the transients restore the micro-causality in eq.(\ref{equal-time}).

\begin{wrapfigure}{r}{0.5\textwidth}
\vspace{-.4in}
\begin{center}
\includegraphics[width=0.5\textwidth,angle=0]{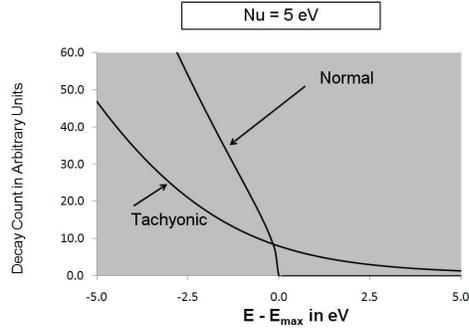}
\label{Spectrum_normvtach}
\vspace{-.25in}
\caption{Spectrum for Normal vs Tachyonic $\nu$}
\end{center}
\vspace{-.3in}
\end{wrapfigure}
\section{Analytic continuation}
But rather than go into the formal aspects of the field theory of tachyons, what I want to show you today is what the true end-point beta spectrum should look like if, indeed, $m_{\nu}^2 < 0$.

When the experimentalist quote their result for a negative $m_{\nu}^2$, they refer to the phenomenological equation (eq.(\ref{decay-rate})), and found that the data fit the equation better if $m_{\nu}^2 = - 1.6 \;eV^2$.  In other words, they simply did an analytic continuation of the rate equation and fitted it.  In this talk, I want to point out in the field theory context what a correct analytic continuation of the rate equation should be.  The result of this exercise is given in the sample figure above and in full analytic form in eq.(\ref{eq-complete spectrum}).  But bear with me as I go into the field theory context.

\section{Survival Amplitude}
\begin{wrapfigure}{r}{0.5\textwidth}
\vspace{-.2in}
\begin{center}
\includegraphics[width=0.5\textwidth,angle=0]{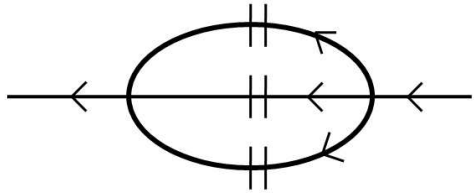}
\label{fig-Cutkosky}
\vspace{-.2in}
\caption{Cutkosky Rule for Survival Amplitude}
\end{center}
\vspace{-.5in}
\end{wrapfigure}
To correctly carry out the analytic continuation, we should go back to the origin of the decay rate formula.  In modern field theory language, we should go back to the Survival Amplitude of the tritium, i.e. the $S-$matrix element for finding the tritium in the out-going state with momentum $P'$ and spin $s'$ given a tritium in the in-coming state
\begin{equation}
    << P', s'; out | P',s; in >> = < P', s'; out| \;T ({\rm e}^{-i \int H_{int} dt}) \;|P', s; in >		\label{survival amplitude}
\end{equation}
\noindent  To second order in weak interactions, we have
\begin{eqnarray}
	\frac{1}{\tau}  \;\delta_{s's} &=& \lim_{T \rightarrow \infty} \frac{1}{T} \;{\cal R}e \; \int_{-T/2}^{+T/2} \int_{-T/2}^{+T/2} dx_o dy_o \int d^3x d^3y  \nonumber \\
									&&\;< P' s'; out | \;
				 T(H_{int} (x) \; H_{int} (y)) \; | P' s; in >  \label{rate eq}
\end{eqnarray}
%	Here $M' -i/\tau$ gives the position of the pole in the tritium propagator.  
and the real part may be obtained by the replacement of all internal lines of momentum $p_i$ according to the Cutkosky rule (Ref ~\refcite{cutkosky}) 
\begin{eqnarray}
	\frac{1}{p_i^2 - m_i^2 + i \epsilon}  &\longrightarrow&  - 2 \pi i \;\delta_{+} (p_{io}^2 - \vec{p}_i \,{}^2 + m_i^2) \;=\; -  \frac{2\pi i}{2\omega} \;\delta (p_{io} - \omega) 		
\end{eqnarray}
This rule comes about when we recognize that it is the positive energy pole of the propagator at $p_{io} =\omega \equiv \sqrt{\vec{p}_i\,{}^2 + m_i^2 }$  that contributes to the $dp_{io}$ integral, 
and for the discontinuity, we compare it with the same loop integral with $m_i^2 + i \epsilon$, so that the discontinuity is given by
\begin{eqnarray}	
	\frac{1}{p_i^2 - m_i^2 + i \epsilon}  \;-\; \frac{1}{p_i^2 - m_i^2 - i \epsilon} &=& -\frac{2 \pi i}{2\omega}\;\delta (p_{io} - \omega)      \label{eq-disc}
\end{eqnarray}
%\noindent with an overall energy-momentum conservation that comes from the integration over space-time.  

\vspace{.1in}
What happens to this rule for a tachyonic neutrino ? 

\vspace{.1in}
By analytic continuation from $m^2 \longrightarrow - {\cal M}^2$ in eq.(\ref{eq-disc}), we expect a new contribution from complex energies associated with the momenta $\vec{q}\;{}^2 < {\cal M}^2$.  
To evaluate this new contribution, we focus on the neutrino propagator in the neighborhood of the positive energy pole and find the discontinuity to be
\begin{eqnarray}
	\frac{1}{q^2 + {\cal M}^2 + i \epsilon} P(q) - \frac{1}{q^2 + {\cal M}^2 - i \epsilon} P(q) &=&
	\;\;\frac{1}{(q_o + i \kappa)(- 2i \kappa )} P(q_o = - i \kappa)  \nonumber  \\
	&& \;-\; \frac{1}{(q_o - i \kappa)( 2i \kappa )} P(q_o = +i\kappa) 
%	&=& \frac{1}{(q_o + i\kappa)(- 2i\kappa)} \label{complex cutkosky}
\end{eqnarray}
Here we have included $P(q)$, a polynomial in the momenta that comes from the spin sum traces, and we have used $\sqrt{\vec{q}\,{}^2 - ({\cal M}^2 \pm i \epsilon)} = \mp i \kappa$, with 
$\kappa \equiv \sqrt{{\cal M}^2 - \vec{q}\,{}^2}$.  And so, even though there are complex energies that lead to exponential run-away modes, we find by ${\cal M}^2 \pm i \epsilon$ analytic continuation that it is the exponentially damped transients that contribute to the decay lifetime.

\section{Tritium $\beta$-decay}

	To investigate these issues, we take the effective interaction Hamiltonian responsible for tritium $\beta$-decay to be given by ($\psi_N$ refers to the tritium field)
\begin{equation}
	H_{int}  \;=\;  \frac{G_F}{\sqrt{2}} \; \overline{\psi}_{_{P}} \gamma_{\mu} (1 + \gamma^5 ) \;\psi_{_N}
						\cdot	\overline{\psi}_{e} \gamma^{\mu} (1 + \gamma^5 ) \;\psi_{\nu}
						+ h.c.
\end{equation}

By inserting the tachyonic propagator\footnote{\samepage It can be verified that the propagator in eq.(\ref{tachyonic propagator}) satisfies the equation of motion in eq.(\ref{pseudo-Dirac}), with the vacuum expectation value of the equal time anti-commutator given by eq.(\ref{equal-time}).
}  
 for the neutrino field(for ease of notation, we have simply replaced ${\cal M}$ by $m_{\nu}$)
\begin{equation}
			< T ( \Psi_{\nu} (x) \overline{\Psi}_{\nu} (y) ) >  =  \frac{1}{i} \int 
			\frac{d^4q}{(2\pi)^4} \;\;\frac{ \gamma \cdot q \gamma^5 + m_{\nu} }{q \cdot q + m_{\nu}^2 + i\epsilon} \;\;{\rm e}^{ -i q \cdot (x - y )}		\label{tachyonic propagator}
\end{equation}
\noindent into eq.(\ref{rate eq}), and performing the space-time integrals, we find that only exponential decay solutions enter in the time-ordered functions, and obtain the result 
\begin{eqnarray}
	\frac{1}{\tau} &=&   \frac{1}{(2P'_o)} \int \;\frac{d^3P}{(2\pi)^3 (2P_o)}
					\frac{d^3p}{(2\pi)^3 (2p_o)}\frac{d^3q}{(2\pi)^3 } \;(2\pi)^3 \delta^{3}(P'-P-p-q) 
					 \nonumber \\
				&&	\left\{ \frac{2 \pi}{2 \omega}\;\delta(P'_o-P_o-p_o-\omega)  
						\;\sum |M|^2\Bigl|_{q_o = \omega} \;\;\theta( \vec{q} \cdot \vec{q} - m_{\nu}^2 )
						\right.\nonumber \\
				&& 				+  \frac{i}{(P'_o - P_o - p_o + i\kappa)(-2i\kappa)} \;\sum |M|^2\Bigl|_{q_o = -i\kappa} 
						\;\;\theta(m_{\nu}^2 - \vec{q} \cdot \vec{q})  \nonumber \\
				&& \left.				-  \frac{i}{(P'_o - P_o - p_o - i\kappa)(\;\;2i\kappa)} 
						\;\sum |M|^2\Bigl|_{q_o = +i\kappa}
						\;\;\theta(m_{\nu}^2 - \vec{q} \cdot \vec{q})			  \right\}  \label{full life-time}
\end{eqnarray}

\noindent	Here the matrix element squared, $\sum |M|^2$, is given by
\begin{eqnarray}
%				\sum | M |^2 &=& 128 \;G_F^2 \;M'^2 \;q_o \; \left( p_o - \frac{m^2}{M'} \right)
				\sum | M |^2 &=& 128 \;G_F^2 \;M'^2 \;q_o \; p_o
\end{eqnarray}
	and $\omega \equiv \sqrt{\vec{q}\cdot\vec{q} - m_{\nu}^2} \ge 0$ is the physical neutrino energy for
	spacelike momenta while $\kappa \equiv \sqrt{m_{\nu}^2 - \vec{q}\cdot\vec{q}}$ is related to the inverse
	lifetime of the transient mode.  
%	$\kappa$ ranges over the values
%\begin{equation}
%	m_{\nu} \;\ge\; \kappa  \ge 0
%\end{equation}

	Eq.(\ref{full life-time}) gives the full contribution to the decay life-time due to the tachyonic pole in
	the propagator ($ q_{\mu} \cdot q^{\mu} + m_{\nu}^2 $).  The energy conserving delta function gives the
	familiar contribution due to physical spacelike neutrino momenta.  The new term arising from the transient
	modes is a Breit-Wigner like enhancement at $x=0$ (see eq.(\ref{eq-x def})).  For very small $m_{\nu}$, this Breit-Wigner term 
	may be approximated by
\begin{equation}
	\frac{2 \kappa}{x^2 + \kappa^2}  \longrightarrow \; 2 \pi \delta (x)
\end{equation}
	However, for small but finite $m_{\nu}$, the transient contribution is not energy conserving, and so there
	is an excess even for $x < 0$, with a width of the order of $m_{\nu}$. (See fig. 6 ).

\section{Complete Decay Spectrum}
	Upon performing the integration over neutrino momenta, the final expression for the lifetime takes the 
	form ($\Delta \equiv M' - M$)
\begin{eqnarray}
	\frac{1}{\tau}  &=&  G_F^2 \cdot \frac{4}{\pi^3} \int_{-|x_{res}|}^{\Delta - m_e} dx \; p\; (\Delta - 
							 x)\left\{  x \cdot \sqrt{m_{\nu}^2 + x^2} \cdot \theta(x)\right. \nonumber \\
				& &	 \left. \;\;\;\;\;\;\;\;\;\;\;\;\;\;\;\;\;\;\;\;\;\;\;\;\;\;\;\;\;\;\;\;\;\;\;\;\;\;\;\;\; \;+\; 
										\frac{m_{\nu}^4}{4\left(|x| + \sqrt{x^2 + m_{\nu}^2}\right)^2} \right\}
										\label{eq-complete spectrum}
\end{eqnarray}
	In this expression, we have introduced the variable, $x$,
\begin{eqnarray}
	x &\equiv& E_{max} - E \;=\; \Delta - E		\label{eq-x def}
\end{eqnarray}
	Note that the lower limit of the $dx$ integration depends on the experimental resolution 
	for $E_{max}$.  Electrons with $x$ below the lower limit,  $ - |x_{res}|$, would have energy $E > E_{max}$ where
\begin{equation}
	E = \Delta  + | x_{res} | \equiv E_{max} + |x_{res}|
\end{equation}
Such electrons are indistinguishable from background non-decay events, and are thus not included in the decay counting rate.

\begin{figure}[tp]
%\begin{center}
%\psfig{file=spectrumtotal.eps,width=5in}
\includegraphics[width=\textwidth,angle=0]{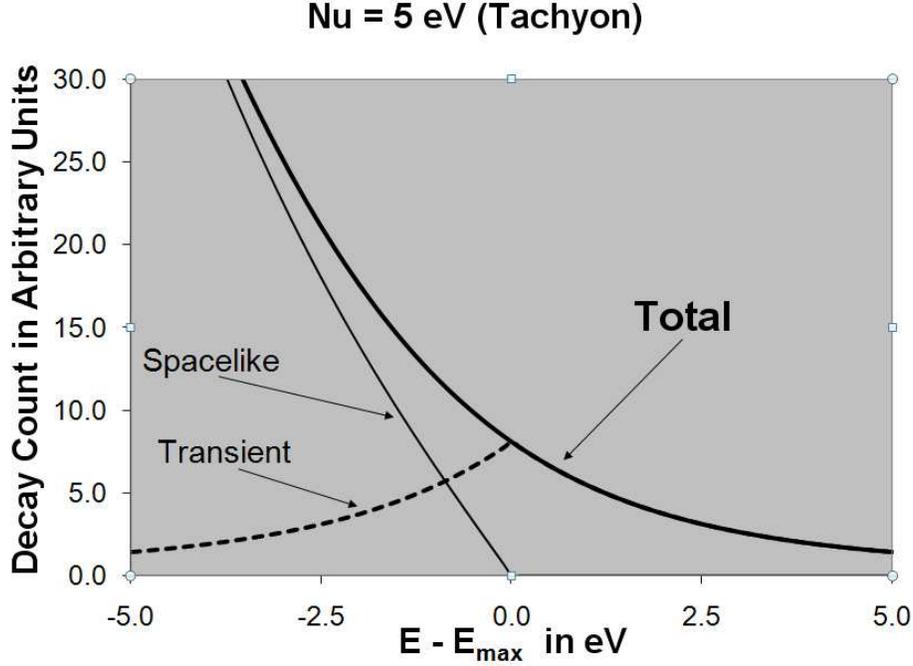}
%\end{center}
\vspace{-.3in}
\label{Spectrum total}
\caption{Total spectrum for Tachyonic Neutrino}
\end{figure}
	
	It may appear surprising if not disconcerting at first sight that the lifetime given in eq.(\ref{rate eq}) should depend on the experimental resolution on the tritium energy.  Where did this dependence on the experimental resolution come from? The answer is that the state-vector, $|P',s'; in >$ is itself an idealization of what is afterall a resonance, and events with electron energies far exceeding $E_{max}$ would be counted as being part of the background rather than as having come from the tritium.
	
	Finally, for completeness, we evaluate the total transient contribution to the tritium lifetime in the $m_{\nu} \rightarrow 0$
	approximation
\begin{equation}
	\frac{1}{\tau} \bigl|_{transient} \;=\; \frac{4 G_F^2 }{3 \pi^2} \;p_{\max} \;\Delta
						\cdot m_{\nu}^3
\end{equation}
	where $p_{max} = \sqrt{\Delta^2 - m_e^2}$.

\section{Conclusion}
The neutrino mass has been an elusive quantity ever since it was conceived by Pauli in 1930, and officially `born' at the Solvay Congress in 1933. That even today we are talking about it is testimony to how elusive the neutrino truly is.  While all the neutrino oscillation data have proven that the neutrinos are massive, the actual nature of the electron neutrino cannot be measured by oscillation alone.  Only the tritium end-point measurements can shed light on its mystery.  The heroic (and largely unsung) efforts by the nuclear physicists in the KATRIN group deserve our respect and our encouragement, for they are forced to forego attention and to be patient as they pursue ever increasing sensitivity in their endpoint measurements. Their data taking is expected to last 5 - 6 years !   

What I have presented here at this PAQFT08 conference is but a small wrinkle to the complete decay spectrum that can help unravel this elusive electron neutrino mass.


\begin{thebibliography}{9}

\bibitem{Troitsk}
	V.M. Lobashev, V.N. Aseev, A.I. Belesev, A.I. Berlev, E.V. Geraskin, A.A. Golubev, 
	O.V. Kazachenko, Yu.E. Kuznetsov, R.P. Ostroumov, L.A. Rivkis, B.E. Stern, N.A. Titov, S.V. Zadorozhny, 
	Yu.I. Zakharov , ``{\em Direct Search for mass of neutrino and Anomaly in the Tritium Beta spectrum}'', 
	\PLB {460}{227}{1999}, \NPProc {91}{280}{2001}.
\bibitem{Mainz}
	Ch. Weinheimer, B. Degen, A. Bleile, J. Bonn, L. Bornschein, O. Kazachenko, A. 
	Kovalik, E.W. Otten, \PLB {460}{219}{1999}, ``{\em High precision measurement of the tritium 
	$\beta$ spectrum near its endpoint and upper limit on the neutrino mass}''.  \\
	For latest update, see
%	``{\em The neutrino mass direct measurements}'', Christian Weinheimer, hep-ex/0306057. 
%	10th Int. Workshop on Neutrino Telescopes, Venice/Italy March 2003. \\
	C. Kraus, J. Bonn, B. Bornschein, L. Bornschein, B. Flatt, A. Kovalik, B. Müller, E.W. Otten, T. Thümmler, J.P. Schall and C. Weinheimer, Eur. Phys. J. C40, 447 (2005), hep-ex/0412056


\bibitem{KATRIN}
	For update on the status of KATRIN, visit {\em http://www-ik.fzk.de/~katrin/index.html} 

\bibitem{tachyon}
	G. Feinberg,
%	LORENTZ INVARIANCE OF TACHYON THEORIES
	\PRD {17} {1651} {1978}. See also
	M.P. Bilaniuk, V.K. Deshpande, E.C.G. Sudarshan, \AmJPhys {30}{718}{1962};
%\bibitem{tachyon-sudarshan}
	J. Dhar, E.C.G. Sudarshan,
	\PhysRev {174} {1808}{1968}.



\bibitem{tachyon-chodos}
	A. Chodos, A. I. Hauser, V.A. Kostelecky, 
	\PLB {150}{431} {1985}.

\bibitem{cutkosky}
	L.D.Landau, \NuclPhys {13}{181} {1959};
	S. Mandelstam, \PhysRev {112}{1344}{1958}, {\em ibid} {\bf 115}, 1741 (1959);
	R.E. Cutkosky, \JMathPhys {1}{429}{1960}. \\
	M. Peskin and D.V. Schroeder, {\em An Introduction to Quantum Field Theory} (Addison-Wesley, 1995).

\bibitem{Chang}
	N.P. Chang, \MPLA {16} {2129} {2001} ({\em hep-ph/0105153})

\end{thebibliography}
\end{document}